\documentstyle{article}
\begin{document}
\title{Quantum probability from a geometrical 
interpretation of a wave function}
\author{Kouji Sugiyama\thanks{
1-4-1-2209,harajuku,Totsuka,Yokohama,Kanagawa,245-0063,JAPAN
E-mail:fwis3728@infoweb.ne.jp}}
\date{November 21, 1999}
\maketitle
\begin{abstract}
The probabilistic prediction of quantum theory is mystery.
I solved the mystery by
a geometrical interpretation of a wave function.
This suggests the unification between quantum theory and
the theory of relativity. 
This suggests Many-Worlds Interpretation is true,too.
\end{abstract} 

\section{Introduction}
There was no theory for Many-Worlds Interpretation 
which can provide the probabilistic prediction of 
quantum theory~\cite{Kent}.
To obtain the probabilistic prediction of quantum theory,
I suppose some assumptions and construct a theory
for Many-Worlds Interpretation.
I call the theory Many-Events Theory.

\section{Assumptions}
\subsection{An assumption of "no boundary"}
There are some physical quantities,
position $x$,time $t$ or phase $\theta$,etc.
I call these physical quantities "position quantity".
I suppose that all position quantities have no boundaries and are compact.

\subsection{An assumption of "a minimum unit"}
I suppose that all position quantities have a common minimum unit $u$.

\subsection{An assumption of "a wave space"}
I suppose new another position quantity $w$. 
I call a space of the quantity a wave space.
This quantity has something to do with a wave function.

\subsection{An assumption of "a phase space"}
I suppose a phase space $\theta$ have a structure
like a M$\ddot{o}$bius strip.
This is analogous with a spin of an electron.
Thus I define a wave function 
\begin{equation}\label{psi}
\psi = R_w exp(\frac{i}{2}\frac{\theta}{R_{\theta}}),
\end{equation}
where $R_w$ is the radius of curvature of a wave space and 
$R_{\theta}$ is the radius of curvature of a phase space.
$R_w$,$R_{\theta}$ and $\theta$ are functions of 
$\theta$,$x$ and $t$.

\subsection{An assumption of "an elementary state"}
I suppose that there is a minimum unit of a state.
I call this an elementary state.
For example in the quantum theory one particle's elementary state is
\begin{equation}\label{state}
|\psi> = |w>|\theta>|x>|t>.
\end{equation}
We can point the elementary state using a combination of position quantities.
I suppose that same elementary state is only one.
Thus we can interpret it as a point in a space.
A state is connected to a set of points.
 
\subsection{An assumption of "an elementary event"}
I call the transition from one elementary state
to another elementary state an elementary event.
I suppose that same elementary event is only one.
Thus we can interpret an elementary event as one line in a space.
An event is connected to a set of lines.
All possible elementary events exist. 
This is analogous with path integrals.

\section{The construction of a torus}
To make the problem easy I think one particle.
And I use $w$,$\theta$,$x$ and $t$.
I construct a torus,
\begin{equation}\label{torus}
F = S^1_w \times _\tau S^1_\theta \times  S^1_x \times S^1_t.
\end{equation}
And I define a point $f$ on the torus,
\begin{equation}\label{point}
f(w,\theta,x,t).
\end{equation}
$R_w$ is linear like a wave function approximately.
If there is a wave on the torus,
an effect of one loop in a phase space and an effect of two loops
cancel out.
Because the phase space has a structure like a M$\ddot{o}$bius strip.
Thus $R_w$ has properties like a wave function.
The shape of the torus is decided by the Hamiltonian $H$ of the system.
To describe many particles, we need a new position quantity.
This is analogous with second quantization.

A point on the torus is an elementary state $f(w,\theta,x,t)$.
We can count elementary states since a position quantity has a minimum unit. 
The number of elementary states $N(\theta,x,t)$ is
\begin{equation}\label{number}
N(\theta,x,t) = \sum_w f(w,\theta,x,t) = \frac{2\pi R_w}{u}
= \frac{2\pi|\psi|}{u}
\end{equation}
where $u$ is a minimum unit and $f(w,\theta,x,t) = 1$.

\section{Probability}
The probability of quantum theory is the probability of an event.
We consider only the number of different elementary events
since we measure only one elementary event of many elementary events.
I suppose that $N(\theta,x,t) = m$ and $N'(\theta',x,t+u) = m' \approx m $.
Then we can obtain the number of different combinations of the
elementary states,$m^2$
since the time has minimum unit.
The combinations are elementary events.
If there are only $m$ different elementary states,
there are $m^2$ different elementary events.

If there is a state $|\Psi> = m|x> + n|x'>$,the probability of
finding a particle at the position $x$,$P(x)$ is 
\begin{equation}\label{probability}
P(x) = \frac{m^2}{m^2+n^2} = \frac{|<x|\Psi>|^2}{<\Psi|\Psi>}.
\end{equation}
Thus we can obtain the probability from this theory.

Each elementary event exists.
Thus each observer measuring each elementary event exists,too.
This is Many-Worlds Interpretation.

\section{Conclusion}
This theory provides the probability of quantum theory.

\end{document}